# Two-photon photoassociation spectroscopy of an ultracold heteronuclear molecule

Sourav Dutta,[1,2,*] Jesús Pérez-Ríos,[2] D. S. Elliott,[2,3,4] and Yong P. Chen[2,3,4]

[1] *Raman Research Institute, C. V. Raman Avenue, Sadashivanagar, Bangalore 560080, India*
[2] *Department of Physics and Astronomy, Purdue University, West Lafayette, IN 47907, USA*
[3] *School of Electrical and Computer Engineering, Purdue University, West Lafayette, IN 47907, USA*
[4] *Purdue Quantum Center, Purdue University, West Lafayette, IN 47907, USA*

We report on two-photon photoassociation (PA) spectroscopy of ultracold heteronuclear LiRb molecules. This is used to determine the binding energies of the loosely bound levels of the electronic ground singlet and the lowest triplet states of LiRb. We observe strong two-photon PA lines with power broadened line widths greater than 20 GHz at relatively low laser intensity of 30 W/cm$^2$. The implication of this observation on direct atom to molecule conversion using stimulated Raman adiabatic passage (STIRAP) is discussed and the prospect for electronic ground state molecule production is theoretically analyzed.

There has been immense interest and progress in creating ultracold heteronuclear polar molecules [1–12] which, by virtue of their permanent electric dipole moment in the electronic ground state, are important for a variety of experiments not possible with ultracold atoms [13–16]. One-photon photoassociation (PA) of ultracold atoms [17] has been extensively used to create ultracold diatomic molecules in excited electronic states [12] and in many cases such molecules spontaneously decay to the electronic ground state [3,18–22]. While this has been a successful strategy even for the formation of rovibronic ground state i.e. $X\ ^1\Sigma^+$ ($v'' = 0$) molecules of some species [3,20,21], the process always leads to molecules being formed in a distribution of vibrational and rotational states [18–22]. An alternative approach is to again start with ultracold atoms but use a coherent coupling scheme for the formation of molecules in a specific rovibronic state. Magnetic Feshbach resonance is one such technique [23] but it can only produce weakly bound molecules and may be of limited use for species with resonances at very high magnetic fields. On the other hand, an all optical scheme such as the Raman-type two-photon PA [24] of ultracold atoms is in principle applicable to all species and is capable of producing ultracold molecules in deeply bound states. The scheme has been previously implemented on homonuclear molecules [25–33] but has never been reported to be observed in heteronuclear bi-alkali molecules.

In this article, we demonstrate Raman-type two-photon PA spectroscopy of the ultracold heteronuclear $^7$Li$^{85}$Rb molecule, a molecule with a relatively large permanent electric dipole moment in its ground state [34], which is promising for a broad array of applications in quantum computing, quantum simulation and ultracold chemistry [13–15]. We also report the binding energies for previously unobserved, loosely bound levels of $^7$Li$^{85}$Rb, compare them with theoretical predictions and derive accurate potential energy curves (PECs) for the $X\ ^1\Sigma^+$ and the $a\ ^3\Sigma^+$ states. In addition, we observe strong two-photon PA resonances with power broadened line widths exceeding 20 GHz. This unexpected feature may suggest new physics not previously encountered and, as we discuss, may lead to efficient formation of ground state molecules.

The experiment is performed in a dual-species magneto-optical trap (MOT) apparatus for simultaneous cooling and trapping of $^7$Li and $^{85}$Rb atoms, the details of which are described elsewhere [35,36]. The $^{85}$Rb MOT is operated as a dark spot MOT in order to reduce atom loss by light assisted collisions [35,36] and to optically pump $^{85}$Rb atoms to the $5s\ ^2S_{1/2}$ ($F = 2$) state. The $^7$Li MOT is a traditional MOT where most of the atoms are in the $2s\ ^2S_{1/2}$ ($F = 2$) state. The scheme used for Raman-type 2-photon PA spectroscopy is shown in Fig. 1(a). The frequency $\nu_{PA}$ of the PA laser is tuned to at a PA transition to create LiRb* molecules in a specific bound vibrational level $v'$ near the Li ($2s\ ^2S_{1/2}$) + Rb ($5p\ ^2P_{1/2}$) atomic asymptote (* indicates electronically excited states). Molecule production leads to a reduction in the number of atoms trapped in the MOT and a corresponding reduction in atomic fluorescence [10,11]. We show an example of this trap loss signal in Fig. 2(a). With $\nu_{PA}$ held fixed on a PA resonance, the frequency $\nu_R$ of a second laser, called the Raman laser, is scanned across a bound-bound $v' \leftrightarrow v''$ transition between the excited and ground electronic states of LiRb. When the frequency $\nu_R$ is resonant with the bound-bound transition, this field strongly couples the two levels and causes shifts (or splitting) in their energies due to a phenomenon commonly known as the Autler-Townes (AT) splitting. Due to the shift in the energy of the $v'$ level of the LiRb* molecule, the PA laser is no longer resonant with the free-bound transition and the atom loss due to PA is suppressed resulting in an increase of the MOT fluorescence [Fig. 2(b)]. The frequency difference $\Delta v\ (= v_R - v_{PA})$ between the Raman and the PA lasers is a measure of the binding energy ($E_B$) of the vibrational level $v''$ of the electronic ground state.

___________
* sourav.dutta.mr@gmail.com , sourav@rri.res.in

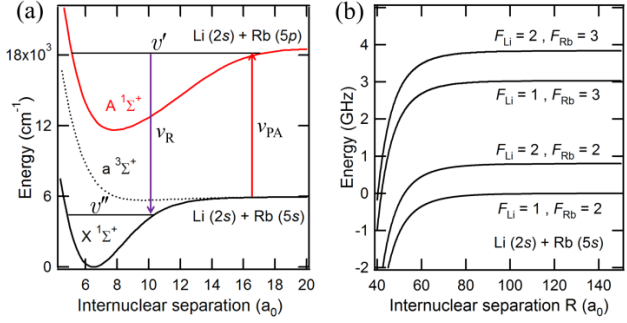

FIG. 1 (Color online). (a) Scheme used for Raman-type 2-photon PA spectroscopy. The frequency $v_{PA}$ of the PA laser is kept fixed while the frequency $v_R$ of the Raman laser is scanned. (b) Electronic ground state PECs at large $R$ [37].

For deeply-bound levels ($\Delta v > 15$ GHz), the signal-to-noise ratio of the Li MOT fluorescence signal is reduced, and we turn to a resonantly-enhanced multi-photon ionization (REMPI) detection scheme [18,19] to detect the 2-photon PA resonances. We use a pulsed dye laser pumped with the second harmonic of a Q-switched Nd:YAG laser to drive the REMPI process. Molecules created by the PA laser spontaneously decay to a distribution of vibrational levels of the $X\ ^1\Sigma^+$ and the $a\ ^3\Sigma^+$ electronic states, which are subsequently ionized using REMPI. In the absence of the Raman laser (or when it is off-resonant), a steady number of LiRb$^+$ ions, typically 2-3 ions per pulse of the dye laser, is detected. When $v_R$ is resonant with the $v' \leftrightarrow v''$ transition, the PA laser frequency is shifted out of resonance due to the AT effect and results in a decrease in ground state molecule formation thereby decreasing the REMPI signal (Fig. 3). The frequencies of the 2-photon PA resonances are not affected by the REMPI laser.

We use a homemade external cavity diode laser (ECDL) [38] as our PA laser with an optical power of ~120 mW and beam diameter of ~ 1 mm. In Fig. 2(a) we show an example of the 1-photon PA spectrum of the $v' = -4$ level of the 2(1) state when the PA laser is scanned in the absence of the Raman laser (negative sign in $v'$ denotes that the level is measured from the dissociation limit). For the Raman-type 2-photon PA spectroscopy, the frequency $v_{PA}$ of the PA laser is fixed (within ±5 MHz) at the center of the PA peak by locking the frequency of the ECDL to a Fabry-Perot cavity. This induces ~20% reduction in the atom number as inferred from the $^7$Li MOT fluorescence. The Raman laser is a Ti:sapphire laser with a maximum output power ~ 400 mW and a beam radius ($r$) of ~ 0.5 mm. It co-propagates along the PA laser albeit with orthogonal linear polarization. The frequency of the Raman laser is scanned in order to obtain the Raman-type 2-photon PA spectra. In Fig. 2(b), we plot the fluorescence of the $^7$Li MOT as a function of $\Delta v\ (= v_R - v_{PA})$. An increase in the $^7$Li MOT fluorescence is seen whenever the Raman laser is resonant with a bound-bound $v' \leftrightarrow v''$ transition.

The decrease in fluorescence at $\Delta v \sim \pm 3$ GHz ($\approx$ ground state hyperfine splitting of $^{85}$Rb) is an artifact and is not related to LiRb molecules. It occurs because the PA and the Raman lasers together transfer, via a Raman transition, the $^{85}$Rb atoms from the lower $F = 2$ to the upper $F = 3$ hyperfine level of the $5s\ ^2S_{1/2}$ state. This disrupts the operation of the dark MOT by putting the $^{85}$Rb atoms in the cycling $5s\ ^2S_{1/2}\ (F = 3) \rightarrow 5p\ ^2P_{3/2}\ (F' = 4)$ transition i.e. essentially turns it to a bright MOT. A bright $^{85}$Rb MOT leads to severe losses in the $^7$Li MOT atom number due to light assisted interspecies collisions as explained in ref. [36]. The decreases in fluorescence at $\Delta v \sim 0, 0.2, 0.5$ and 1.7 GHz are due to 1-photon PA transitions induced by the Raman laser – the first three corresponds to the 2(1), $v' = -4$ level, while the last corresponds to the 2(0$^-$), $v' = -4$ level [10].

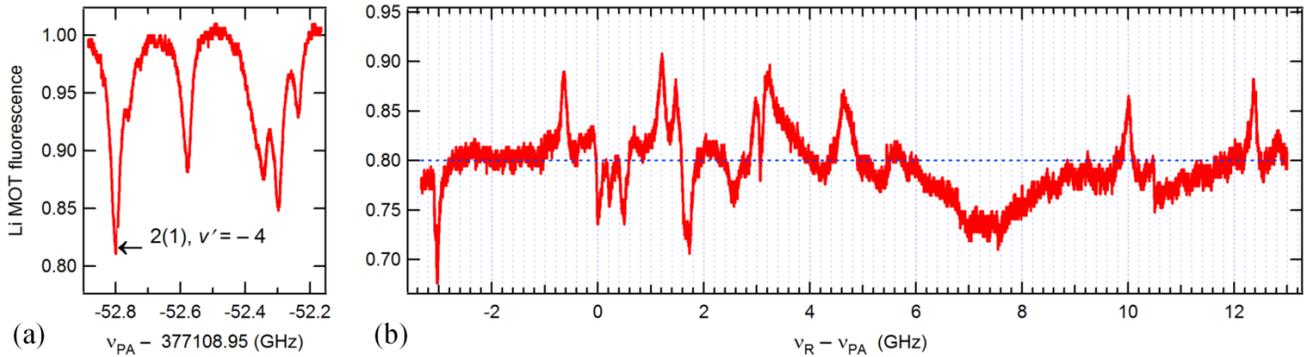

FIG. 2 (Color online). (a) A typical 1-photon PA spectrum. (b) A 2-photon PA spectrum when the frequency of the Raman laser is scanned, with the ECDL locked to the 2(1), $v' = -4$ PA line indicated by the arrow in (a).

Table I. Binding energies $E_B$ (in GHz) of experimentally observed 2-photon PA resonances (in the experimentally accessed spectral range $3.5 \leq \Delta \nu \leq 200$ GHz) and their assignments. Also tabulated are the values of $E_B$ calculated using the PECs of ref. [39] and those derived in this work (also see Supplementary Material [40]).

| $E_B (= \Delta \nu)$ Experiment | Assignment State ($v''$, $N''$) | $E_B$ Ref. [39] | $E_B$ This work |
|---|---|---|---|
| 4.51(9) | $a\,^3\Sigma^+$ (14, 0) | 3.93 | 4.66 |
| 10.03(3) | $X\,^1\Sigma^+$ (51, 2) | 11.56 | 9.80 |
| 12.42(4) | $X\,^1\Sigma^+$ (51, 0) | 13.93 | 12.19 |
| 32.2(4) | $a\,^3\Sigma^+$ (13, 0) | 33.01 | 31.21 |
| 54.7(3) | $X\,^1\Sigma^+$ (50, 0) | 67.51 | 54.96 |
| 100.5(3) | $a\,^3\Sigma^+$ (12, 0) | 111.51 | 100.92 |

We now discuss assignments for some of the observed 2-photon PA spectra of $^7$Li$^{85}$Rb. The levels bound by less than 2 GHz were assigned using a recently developed method based on quantum defect theory (QDT) as described elsewhere [41]. Here, we focus on levels with binding energies greater than 2 GHz which, being more deeply bound, are better intermediates for transfer of molecules to the $v'' = 0$ level of the $X\,^1\Sigma^+$ and the $a\,^3\Sigma^+$ electronic states. In the Raman-type 2-photon PA spectra, only the rotational levels $N'' = 0$ and $N'' = 2$ are visible, where $N''$ is the nuclear rotational quantum number for the bound ground state molecule. This happens because the two-photon process couples states of same symmetry leading to the selection rule $\Delta N = (N'' - N_{sc}) = 0, \pm 2$, where $N_{sc}$ is the nuclear rotational quantum number for the scattering state. In our case $N_{sc} = 0$ since primarily $s$-wave scattering is possible at the sub mK temperatures of the MOTs [10,11], restricting $N''$ to 0 or 2. The observation (see Table I) of the $N'' = 0$ and $N'' = 2$ rotational levels of the $X\,^1\Sigma^+$ ($v'' = 51$) state confirms the selection rule and the predominance of $s$-wave collisions. Further, the frequency spacing between $N'' = 0$ and $N'' = 2$ rotational levels agrees very well with calculations [39,40]. Also note that the $X\,^1\Sigma^+$ and the $a\,^3\Sigma^+$ potentials have four dissociation asymptotes – these correspond to the Li (2$s$, $F = 1$) + Rb (5$s$, $F = 2$), Li (2$s$, $F = 2$) + Rb (5$s$, $F = 2$), Li (2$s$, $F = 1$) + Rb (5$s$, $F = 3$) and Li (2$s$, $F = 2$) + Rb (5$s$, $F = 3$) channels. The differences in the energy of these channels correspond to the atomic hyperfine splitting, as shown in Fig. 1(b). In our experiment the free Li and Rb atoms collide mainly in the Li (2$s$, $F = 2$) + Rb (5$s$, $F = 2$) channel, so we quote binding energies with respect to this asymptote. Whether an observed line belongs to the $a\,^3\Sigma^+$ state or to the $X\,^1\Sigma^+$ state is determined by examining the hyperfine structure. The levels belonging to the $X\,^1\Sigma^+$ state are not expected to have any hyperfine structure while those belonging to the $a\,^3\Sigma^+$ state are. As shown in Fig. 3, each $v''$ level of the $a\,^3\Sigma^+$ state splits into three hyperfine levels. This occurs because the total spin ($S = 1$) of this triplet state adds with the nuclear spin ($I_{Rb} = 5/2$) of $^{85}$Rb, resulting in three values of $\mathbf{G}$ (= $\mathbf{I_{Rb}} + \mathbf{S}$). The frequency span of the hyperfine structure, as expected, is observed to be of the same order as the hyperfine splitting (3.04 GHz) of the $^{85}$Rb atoms in the 5$s$ $^2S_{1/2}$ state. In principle, each of these lines should split further due to coupling with the nuclear spin of $^7$Li but this is small (hyperfine splitting of $^7$Li is 0.8 GHz) and not clearly resolved in our measurement. Once the electronic state for an observed line is determined, we assign the $v''$ level by comparing the observed line positions (i.e. the binding energies) with those calculated using the PECs for the $X\,^1\Sigma^+$ and $a\,^3\Sigma^+$ states in ref. [39]. The $X\,^1\Sigma^+$ and $a\,^3\Sigma^+$ states have 53 ($v'' = 0$-52) and 15 ($v'' = 0$-14) vibrational states, respectively. The binding energies calculated using the PECs in ref. [39] are in reasonable agreement with experimentally observed values (see Table I) but deviate significantly in some cases. We thus calculate new PECs (see supplementary material [40] for details) which predict binding energies in much better agreement with the experiment (see Table I). The derived PECs reproduce all observed vibrational levels of the $X\,^1\Sigma^+$ and $a\,^3\Sigma^+$ states with an accuracy of 0.02 cm$^{-1}$ and 0.3 cm$^{-1}$, respectively. We note that the finer structures in the observed lines are poorly resolved and yet to be assigned. Similarly, some expected states, such as the $X\,^1\Sigma^+$ ($v'' = 49$), were not observed experimentally – plausible reasons for which are given below.

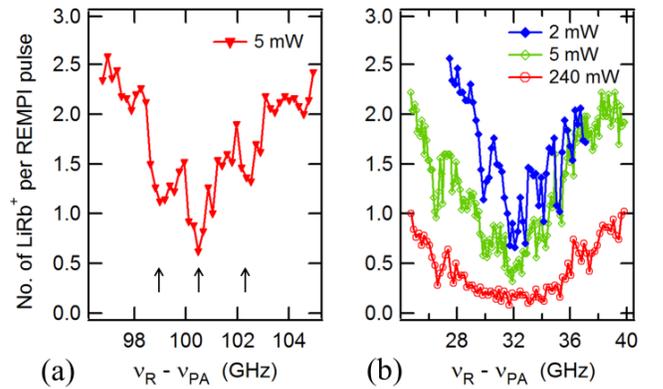

FIG. 3 (Color online). Two-photon PA spectra measured using the REMPI detection scheme. The PA laser is tuned to the 2(1) ($v' = -5$) level and the REMPI laser (frequency 17871.5 cm$^{-1}$) ionizes molecules produced in the $a\,^3\Sigma^+$ ($v'' = 13$) state by spontaneous emission. (a) The 2-photon PA resonance corresponding to the $a\,^3\Sigma^+$ ($v'' = 12$) ↔ 2(1) ($v' = -5$) transition. The triplet structure due to hyperfine splitting of the $a\,^3\Sigma^+$ ($v'' = 12$) level is indicated by arrows. (b) The $a\,^3\Sigma^+$ ($v'' = 13$) ↔ 2(1) ($v' = -5$) transition measured using different Raman laser powers as indicated in the legends.

We repeated the 2-photon PA measurements for different 1-photon PA transitions. This was done because of the following reasons: (*i*) to verify that the 2-photon PA resonances, or at least some of them, can be reproduced for other 1-photon PA transitions, (*ii*) to verify that the Raman laser couples the photoassociated level to the ground electronic state and not to a more highly excited electronic state (in case of the latter, the 2-photon resonances would not be observed at the same $\Delta\nu$ on choosing a different 1-photon PA transition), (*iii*) the spin singlet $X\ ^1\Sigma^+$ state and the spin triplet $a\ ^3\Sigma^+$ state, in absence of singlet-triplet mixing, can couple only to singlet and triplet PA levels, respectively. Thus, in order to observe both the $X\ ^1\Sigma^+$ and the $a\ ^3\Sigma^+$ states, different excited states need to be chosen (in reality, some singlet-triplet mixing does exist making this selection rule somewhat relaxed), and (*iv*) different $v'$ levels have outer turning points at different internuclear separations (~ 30-45 $a_0$ for the states considered here) and the Franck-Condon (FC) overlap with the $v''$ of the ground electronic state will be different for different $v'$ levels. It is expected that more deeply bound PA levels will have better FC overlap with more deeply bound levels of the electronic ground state. This limits the number of $v''$ levels that can be observed for a particular $v'$ level.

We note that we observe 2-photon PA to both $a\ ^3\Sigma^+$ and $X\ ^1\Sigma^+$ states irrespective of whether the 1-photon PA laser is tuned to the 2(1) ($v'$ = -4, -5, -6), 2(0$^-$) ($v'$ = -5) states which have spin triplet character or to the 2(0$^+$) ($v'$ = -4) level that has spin singlet character. This suggests the presence of the expected singlet-triplet mixing that has so far been crucial [2] for the production of molecules in the $X\ ^1\Sigma^+$ ($v''$ = 0) level.

It is also interesting to note that some of the observed lines have very broad line widths. Figure 3(b) shows the $a\ ^3\Sigma^+$ ($v''$ = 13) ↔ 2(1) ($v'$ = -5) transition measured using different intensities of the Raman laser. A Lorentzian fit to the central hyperfine feature yields FWHM $\delta$ = 3.0(3), 5.1(4), 21.7(6) GHz for Raman laser powers $P$ = 2, 5, 240 mW, respectively. A fit of the $\delta$ vs. $P$ data to a simple model where $\delta = \delta_0(1 + P/P_{sat})^{1/2}$ [42] yields $\delta_0 \approx$ 3.0(9) GHz and saturation power $P_{sat} \approx$ 0.6(3) mW. The saturation intensity is $I_{sat}(= P_{sat}/\pi r^2) \approx$ 0.08(4) Wcm$^{-2}$ – this unprecedented low saturation intensity indicates that the $a\ ^3\Sigma^+$ ($v''$ = 13) ↔ 2(1) ($v'$ = -5) transition is extremely strong. In an earlier report we had shown that the PA rate to 2(1) ($v'$ = -5) is also quite high at relatively low PA laser intensity [10,11]. Together these allow the possibility of direct conversion of free $^7$Li and $^{85}$Rb atoms to bound $^7$Li$^{85}$Rb molecules in the $a\ ^3\Sigma^+$ ($v''$ = 13) using stimulated Raman adiabatic passage (STIRAP). Such all optical conversion of atoms to molecules can be extremely efficient and fast, and may offer a pathway to create high density samples of quantum degenerate molecules [43–47].

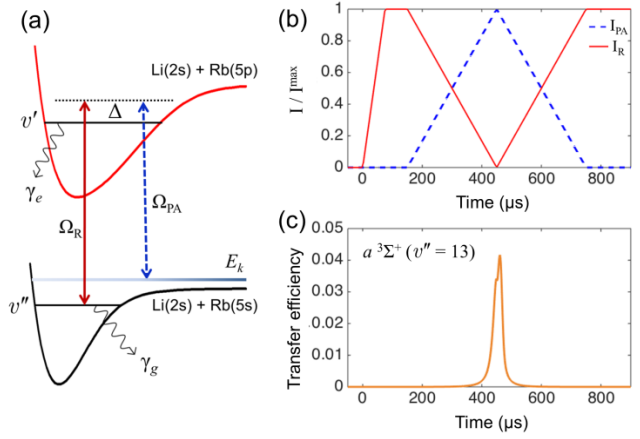

FIG. 4 (Color online). (a) Scheme for free-to-bound STIRAP (see text for details). (b) Laser pulse sequence employed in the simulations for STIRAP. (c) Solution of the time-dependent Schrödinger equation for the proposed STIRAP scheme, for the case when the final state of the molecules is the $a\ ^3\Sigma^+$ ($v''$ = 13) level.

In order to explore such a possibility theoretically, we assume an optical dipole trap loaded with atoms at a temperature of 50 μK and atomic density $\rho = 10^{12}$ cm$^{-3}$. The present STIRAP scheme is reminiscent of the bound-bound STIRAP technique [48], as shown in Fig. 4(a), but with the subtleties of the free-to-bound transition included. We consider transfer to $a\ ^3\Sigma^+$ ($v''$ = 13) level using 2(1) ($v'$ = -5) as the intermediate level. The PA transition is characterized by the coupling rate $\Omega_{PA} = \langle \varphi_{E_k} | \varphi_{v'=-5} \rangle \Omega_{at} \sqrt{I_{PA}\rho/I_s\hbar^2}$ [49], where $\varphi_{E_k}$ and $\varphi_{v'=-5}$ represent the energy-normalized continuum wave function and the excited state wave function, respectively. $I_{PA}$ ($I_R$) is the intensity of the PA (Raman) laser, $I_s$ is the saturation intensity for the Rb D$_1$ line and $\Omega_{at} = \Gamma/\sqrt{8}$ [50], $\Gamma$ being the natural line width of the Rb D$_1$ line. It is assumed that the PA occurs in the long-range region of atom-atom interaction [50], otherwise the molecular transition dipole moment should be included in between the bracket associated with the free-to-bound overlap. $\Omega_R$ is the Rabi frequency for the bound-bound transition. The decay rates of the excited and ground state levels are assumed to be $\gamma_e = 2\pi \times 5.75$ MHz and $\gamma_g = 2\pi \times$ 3 kHz, respectively [48]. We also include the detuning $\Delta = 2\pi \times 0.8$ MHz due to the AC Stark shift induced by the laser fields. Figure 4(b) shows the optimum time sequence of the lasers used for STIRAP. For the calculations we use $I_{PA}^{max}$ = $10^4$ Wcm$^{-2}$ and $I_R^{max}$ = 1 Wcm$^{-2}$ leading to $\Omega_{PA}^{max} = 2\pi \times 94.9$ kHz and $\Omega_R^{max} = 2\pi \times 10.67$ MHz, respectively. The time-dependent Schrödinger equation is solved for the effective three-level system after applying the rotating wave approximation, and the results are shown in Fig. 4(c). We find the free to bound transfer efficiency to be 4%. Although not very high, this leads to $4 \times 10^4$ molecules in the $a\ ^3\Sigma^+$ ($v'' = 13$) level, starting with ~$10^6$ ultracold atoms that are routinely available in experiments. We also calculated the efficiency for transfer to the $X\ ^1\Sigma^+$ ($v'' = 43$) level which has a binding energy of 137 cm$^{-1}$ and a permanent electric dipole moment of ~1.5 Debye [34]. With the same laser powers and pulse sequence, but using as intermediate the $B\ ^1\Pi$ ($v' = 20$) level [35,51], located 125 GHz below the $^7$Li ($2s\ ^2$S$_{1/2}$) + $^{85}$Rb ($5p\ ^2$P$_{1/2}$) asymptote [22], we find a transfer efficiency of 1.7% that corresponds to $1.7 \times 10^4$ molecules in a single rovibrational $X\ ^1\Sigma^+$ ($v'' = 43$) level, starting with ~$10^6$ atoms. Such high efficiency is encouraging for all-optical production of ultracold molecules.

In summary, we demonstrate two-photon PA spectroscopy of a heteronuclear bi-alkali molecule and use it to determine the binding energies for the previously unobserved, loosely bound levels of $X\ ^1\Sigma^+$ and $a\ ^3\Sigma^+$ states of the $^7$Li$^{85}$Rb molecule. We observe strong two-photon PA lines and discuss the implication of this result in the formation of copious amounts of ground state $^7$Li$^{85}$Rb molecules. Future experiments, with atoms loaded in an optical dipole trap, could enable efficient production of rovibronic ground state LiRb molecules.

S.D. acknowledges partial support from Department of Science and Technology (DST), India in form of the DST-INSPIRE Faculty Award (Award No. IFA14-PH-114).

---

## Supplementary Material

**Potential energy curve and the energy levels of the $X\ ^1\Sigma^+$ state of LiRb ($^7$Li$^{85}$Rb) molecule:**

The potential energy curve and the energy levels are obtained using LeRoy's RKR code [52] and using as inputs the experimentally observed line positions. The v″ = 0 - 45 lines were observed using laser induced fluorescence and v″ = 50, 51 are observed using 2-photon photoassociation spectroscopy. Other vibrational levels have not been experimentally observed yet. The maximum discrepancy between the calculated energy levels (given below) and the experimentally observed line positions is 0.02 cm$^{-1}$. For each vibrational level v″: the rotationless energy $E$(v″), the rotational constant $B$(v″) are shown, along with the inner turning point $R_{\min}$(v″) and the outer turning point $R_{\max}$(v″). The dissociation energy of $^7$Li$^{85}$Rb is $D_e$ = 5928.11(2) cm$^{-1}$ with respect to the [$^{85}$Rb 5$s$ ($F$ = 2) + $^7$Li 2$s$ ($F$ = 2)] dissociation asymptote. The numbers below certainly contain more digits than necessary.

| v″ | $E$(v″) in cm$^{-1}$ | $B$(v″) in cm$^{-1}$ | $R_{\min}$(v″) in Å | $R_{\max}$(v″) in Å |
|---|---|---|---|---|
| 0  | 97.2793   | 0.215705504 | 3.3109927224 | 3.6383316592 |
| 1  | 290.2399  | 0.214258116 | 3.2053217069 | 3.7749396221 |
| 2  | 480.978   | 0.212736224 | 3.1366229421 | 3.8755931855 |
| 3  | 669.4454  | 0.211163483 | 3.0831385086 | 3.9618836175 |
| 4  | 855.6233  | 0.209556049 | 3.0385354873 | 4.0400228663 |
| 5  | 1039.4983 | 0.207924016 | 2.9999083917 | 4.1128376678 |
| 6  | 1221.0536 | 0.206272685 | 2.9656360851 | 4.1819068008 |
| 7  | 1400.2668 | 0.204603663 | 2.9347077641 | 4.2482286364 |
| 8  | 1577.1108 | 0.202915817 | 2.9064472731 | 4.3124907183 |
| 9  | 1751.5558 | 0.201206075 | 2.8803801577 | 4.3751974523 |
| 10 | 1923.5701 | 0.199470103 | 2.8561616036 | 4.4367379083 |
| 11 | 2093.1207 | 0.197702842 | 2.8335338333 | 4.4974251244 |
| 12 | 2260.1728 | 0.195898942 | 2.8122992568 | 4.5575204831 |
| 13 | 2424.6896 | 0.194053077 | 2.7923027792 | 4.6172496067 |
| 14 | 2586.631  | 0.192160174 | 2.7734198044 | 4.6768130983 |
| 15 | 2745.9532 | 0.190215538 | 2.7555479682 | 4.7363940070 |
| 16 | 2902.6078 | 0.188214913 | 2.7386013966 | 4.7961631776 |
| 17 | 3056.5423 | 0.18615446 | 2.7225067017 | 4.8562832745 |
| 18 | 3207.6993 | 0.18403068 | 2.7072001783 | 4.9169120524 |
| 19 | 3356.0169 | 0.181840281 | 2.6926258243 | 4.9782053059 |
| 20 | 3501.4289 | 0.179580007 | 2.6787339257 | 5.0403198186 |
| 21 | 3643.864 | 0.17724643 | 2.6654800323 | 5.1034165464 |
| 22 | 3783.2466 | 0.174835713 | 2.6528242144 | 5.1676641860 |
| 23 | 3919.4953 | 0.17234337 | 2.6407305435 | 5.2332432197 |
| 24 | 4052.523 | 0.169764005 | 2.6291667678 | 5.3003504817 |
| 25 | 4182.2358 | 0.167091066 | 2.6181041770 | 5.3692042721 |
| 26 | 4308.5325 | 0.164316602 | 2.6075176530 | 5.4400500582 |
| 27 | 4431.3035 | 0.161431052 | 2.5973858954 | 5.5131668531 |
| 28 | 4550.4307 | 0.158423052 | 2.5876917955 | 5.5888744546 |
| 29 | 4665.7868 | 0.155279292 | 2.5784229064 | 5.6675418734 |
| 30 | 4777.2353 | 0.151984419 | 2.5695719286 | 5.7495974761 |
| 31 | 4884.6303 | 0.148520995 | 2.5611373280 | 5.8355418637 |
| 32 | 4987.8166 | 0.144869528 | 2.5531187251 | 5.9259593990 |
| 33 | 5086.6295 | 0.141008575 | 2.5455168683 | 6.0215399549 |
| 34 | 5180.8948 | 0.136914934 | 2.5383336261 | 6.1231087084 |
| 35 | 5270.428 | 0.13256393 | 2.5315720001 | 6.2316672439 |
| 36 | 5355.0342 | 0.127929803 | 2.5252361526 | 6.3484504841 |
| 37 | 5434.5077 | 0.122986217 | 2.5193314309 | 6.4750058876 |
| 38 | 5508.6318 | 0.117706881 | 2.5138643566 | 6.6133044349 |
| 39 | 5577.1809 | 0.11206631 | 2.5088425286 | 6.7658980981 |
| 40 | 5639.9237 | 0.106040727 | 2.5042743731 | 6.9361474845 |
| 41 | 5696.631 | 0.099609108 | 2.5001686555 | 7.1285595474 |
| 42 | 5747.0878 | 0.092754395 | 2.4965336552 | 7.3493055378 |
| 43 | 5791.1123 | 0.085464876 | 2.4933759052 | 7.6070485022 |
| 44 | 5828.5818 | 0.077735737 | 2.4906984063 | 7.9143314309 |
| 45 | 5859.468 | 0.069570809 | 2.4884982681 | 8.2900451410 |
| 46 | 5883.8788 | 0.060984506 | 2.4867638008 | 8.7641348087 |
| 47 | 5902.106 | 0.052003973 | 2.4854712124 | 9.3873960630 |
| 48 | 5914.6732 | 0.042671444 | 2.4845812650 | 10.2543091532 |
| 49 | 5922.3724 | 0.033046828 | 2.4840365498 | 11.5650635771 |
| 50 | 5926.2766 | 0.023210528 | 2.4837604692 | 13.8337157801 |
| 51 | 5927.7034 | 0.013266496 | 2.4836596032 | 18.7425832259 |
| 52 | 5928.0951 | 0.003345553 | 2.4836319132 | 26.3801898311 |

To extrapolate the potential to $R < 2.4836319132$ Å, use the expression:
$$V(R) = -5050.0195 + 269444.73 \, exp(-1.28862073 \, R).$$

To extrapolate the potential to $R > 26.3801898311$ Å, one could use the long range expression:
$V(R) = 5928.11 - C_6/R^6 - C_8/R^8 - C_{10}/R^{10}$,
where $C_6$, $C_8$ and $C_{10}$ recommended by ref. [53] are:
$C_6 = 1.228961D+07$ cm$^{-1}$Å$^6$ (*i.e.* 2550.04 a.u.),
$C_8 = 3.171570D+08$ cm$^{-1}$Å$^8$ (*i.e.* 235007.6 a.u.),
$C_{10} = 1.295324D+10$ cm$^{-1}$Å$^{10}$ (*i.e.* 34275526 a.u.).

*Note*:

The value of $C_6$ given above agrees well with $C_6$ = 2545(7) a.u. calculated in ref. [37].

The value of $C_8$ given above agrees well with $C_8$ = 2.34D+05 a.u. calculated in ref. [54].

However, the value of $C_{10}$ given above differs from $C_{10}$ = 2.61D+07 a.u. calculated in ref. [54].

**Potential energy curve and the energy levels of the $a\ ^3\Sigma^+$ state of LiRb ($^7$Li$^{85}$Rb) molecule:**

The potential energy curve and the energy levels are obtained using LeRoy's RKR code [52] and using as inputs the experimentally observed line positions. The v″ = 1, 2, 4-13 lines are observed using REMPI and v″ = 12-14 are observed using 2-photon photoassociation spectroscopy. Vibrational levels v″ = 0, 3 have not been experimentally observed yet. The maximum discrepancy between the calculated energy levels (given below) and the experimentally observed REMPI line positions is 0.3 cm$^{-1}$ (consistent with the error in REMPI spectra). The calculated position of lines v″ = 12-14 agree with 2-photon PA line positions within 0.04 cm$^{-1}$ (consistent with the error in 2-photon PA line position arising from the molecular hyperfine splitting of ~0.1 cm$^{-1}$). For each vibrational level v″: the rotationless energy $E$(v″), the rotational constant $B$(v″) are shown, along with the inner turning point $R_{min}$(v″) and the outer turning point $R_{max}$(v″). The dissociation energy of $^7$Li$^{85}$Rb is $D_e$ = 278.77(30) cm$^{-1}$ with respect to the [$^{85}$Rb 5$s$ ($F$ = 2) + $^7$Li 2$s$ ($F$ = 2)] dissociation asymptote. The numbers below certainly contain more digits than necessary.

| v″ | $E$(v″) in cm$^{-1}$ | $B$(v″) in cm$^{-1}$ | $R_{min}$(v″) in Å | $R_{max}$(v″) in Å |
|---|---|---|---|---|
| 0 | 19.9491 | 0.096423027 | 4.8235715345 | 5.5496545455 |
| 1 | 57.7711 | 0.091986509 | 4.6463898002 | 5.9440584030 |
| 2 | 92.616 | 0.087587665 | 4.5428555009 | 6.2777663770 |
| 3 | 124.5101 | 0.083122623 | 4.4689711792 | 6.5994379196 |
| 4 | 153.4952 | 0.078496505 | 4.4126762780 | 6.9271633720 |
| 5 | 179.5248 | 0.073623427 | 4.3682270470 | 7.2750859835 |
| 6 | 202.4945 | 0.068426501 | 4.3322918818 | 7.6579665277 |
| 7 | 222.3145 | 0.062837833 | 4.3033991411 | 8.0931794483 |
| 8 | 238.964 | 0.056798523 | 4.280461307 | 8.6017663800 |
| 9 | 252.4985 | 0.050258667 | 4.2626256621 | 9.2141126786 |
| 10 | 263.0146 | 0.043177355 | 4.2492314816 | 9.9848010452 |
| 11 | 270.6095 | 0.035522671 | 4.2397957547 | 11.0322742300 |
| 12 | 275.4037 | 0.027271696 | 4.2339380701 | 12.6605648951 |
| 13 | 277.7288 | 0.018410502 | 4.2311240174 | 15.7974008505 |
| 14 | 278.6144 | 0.00893416 | 4.2300565811 | 20.1712123549 |

To extrapolate the potential to $R$ < 4.2300565811 Å, use the expression:
$V(R)$ = -102.0575 + 3885218.8 $exp$(-2.18218166 $R$).